\documentclass[12pt]{article}

\newcommand{\lanln}[1]{$\langle$\texttt{arXiv:#1}$\rangle$}
\usepackage{verbatim}

\usepackage{color}
\usepackage{amsfonts}
\usepackage{amsmath}
\usepackage{amssymb}
\usepackage{latexsym}
\usepackage{psfrag}
\usepackage{lastpage}

\usepackage[pdftex]{graphicx} 
\usepackage[center]{caption2}

\usepackage{hyperref}

\date{}
\begin{document}

\begin{center} {\Large \bf
Black hole entanglement entropy and the\\
\vskip 2mm
 renormalization group}
\end{center}

\vskip 5mm
\begin{center} \large

{{Ted Jacobson\footnote{jacobson@umd.edu} and Alejandro Satz\footnote{alesatz@umd.edu}}}
\end{center}
\vskip  0.3 cm
{\centerline{\it Maryland Center for Fundamental Physics}}
{\centerline{\it Department of Physics, University of Maryland}}

{\centerline{\it College Park, MD 20742-4111, USA}}

\vskip  0.4 cm
{\centerline{\it Kavli Institute for Theoretical Physics}}
{\centerline{\it University of California}}

{\centerline{\it Santa Barbara, CA 93106-4030, USA}}

\vskip 1cm

\begin{abstract}\small
{
We investigate the contributions of quantum fields to black hole entropy by using a cutoff scale at which the theory is described with a Wilsonian effective action. For both free and interacting fields, the total black hole entropy can be partitioned into a contribution derived from the gravitational effective action and a contribution from quantum fluctuations below the cutoff scale. In general the latter includes a quantum contribution to the Noether charge. We analyze whether it is appropriate to identify the rest with horizon entanglement entropy,  and find several complications for this interpretation,
which are especially problematic for interacting fields.
}
\end{abstract}

\section{Introduction}
The concept of black hole entropy, first introduced by Bekenstein \cite{beken1, beken2} 
using information theory and the
analogy between black hole mechanics and thermodynamics, became firmly established when Hawking \cite{hawk1} derived black hole radiation and its thermal properties by considering 
quantum fields on a black hole background.
The formula expressing the black hole entropy in terms of the horizon area $A$ and Newton's constant,
\begin{equation}
S_{BH}=\frac{A_{}}{4\hbar G/c^3},
\end{equation}
has since then become a 
focal point 
of quantum gravity research.
Attempts to derive it from an analysis of microscopic, fundamental degrees of freedom
have met with varying degrees of success in string theory, loop quantum gravity, and other approaches.

Staying within an effective treatment in which gravity is described by a metric field, there appear to be two contributions to black hole entropy. Firstly, the gravitational field itself in the absence of matter fields seems to have a 
``gravitational"
entropy given by the Bekenstein-Hawking formula. 
This can be derived from the first law of black hole mechanics and the Hawking temperature, or
from the saddle-point evaluation of the Euclidean gravitational path integral 
for the thermal partition function\cite{gibbonshawk}.
Secondly, 
the contribution of 
quantum matter fields on the black hole background  
(and of gravitons, if metric fluctuations are quantized perturbatively) 
to the thermal entropy 
is also proportional, in the leading order, to the event horizon area, but with a divergent coefficient. This entropy 
arises from the one-loop correction to the thermal partition function. It 
can be conceptualized 
as the entropy of a thermal state for quantum fields outside the horizon, or,
at least in some cases,  
as the entanglement entropy across the horizon of quantum fields in a global pure state. 
For reviews of the main issues and results, see \cite{frolov1, solod1}.

The area-scaled divergence in the matter contribution to the entropy can be absorbed into a renormalization of Newton's constant. More precisely, the divergences in the entropy are related to the ``bare gravitational'' entropy in the same way as the divergences in the effective gravitational action are related to the bare gravitational action \cite{suss}. The renormalization properties of the black hole entanglement entropy have been further studied in \cite{PV, larsen, jacobsonren},  with the 
case of non-minimally coupled fields receiving special attention both for scalars \cite{solod3} and for gauge fields \cite{kabat, donn, solod4}.  
Different regularization methods
(brick-wall boundary at the horizon, Pauli-Villars regulator, UV cutoff $\epsilon$ in heat kernel expansion...) do not always give the same results for the matter contribution\cite{PV, thooft, reg1,reg2,reg3, dimen}. These discrepancies are not surprising, since the UV regulator modifies precisely those degrees of freedom which are most responsible for the entropy.

While black hole entropy emerges, formally, from the gravitational partition function, 
that does not reveal the nature of the states that are counted by the entropy. 
It is tempting to think that, like the 
minimally
coupled matter contribution, all of the entropy 
might be interpreted as entanglement entropy. For this to make sense, it would seem
that the low energy Newton constant must arise fully from integrating out 
quantum fluctuations,
i.e.\ that there is no ``bare" gravitational action or entropy at the UV cutoff scale, so that gravity
is entirely ``induced"\cite{jacobsonren}. 
From the QFT viewpoint there seems to be no reason why this should be so,
although a thermodynamic argument suggests it must be \cite{TOS, ENT}.  But, in any case,
without a UV completion of the theory, it is not really possible to 
assess
this entanglement interpretation of the full entropy because the value of the entropy depends on the 
artificial UV cutoff at scale $\Lambda_{UV}$. 

However, it is possible to test the entanglement interpretation of black hole
entropy in a way that side-steps the unkown physics of the UV cutoff 
and deals only with finite quantities.
We can partition the degrees of freedom into those with momenta greater than an
intermediate scale $k\ll \Lambda_{UV}$, which are integrated out 
and absorbed into a Wilsonian effective action, 
and those with momenta 
less than $k$, whose quantum fluctuations contribute explicitly to the entropy. 
As $k$ is lowered, contributions to the total entropy transfer from the
explicit fluctuations to the ``gravitational entropy" of the effective action,
i.e.\ the area term plus curvature corrections,
via the flow of the gravitational couplings. 
This allows us to exhibit the renormalization behavior of the contributions 
to the entropy without sensitivity to the UV cutoff, which in turn allows us to 
make precise sense of the question whether the entropy of the modes with
momenta less than $k$ admits an interpretation
as entanglement entropy. We shall find that such an interpretation, 
while at first superficially plausible, suffers from a number of difficulties.

The renormalization group (RG)
flow of black hole entropy from state counting of explicit field fluctuations
to effective gravitational action was studied long ago in 
Ref.~\cite{onmodel}.
That paper studied
how the entropy accounting changes when the renormalization scale
drops below a mass scale of the fluctuations, and  focused on the interpretation of the 
contribution from non-minimal couplings to curvature. Our study is very similar in 
spirit, but we consider continuously varying RG scale, and we try to assess 
more precisely the validity of the state counting interpretation
even when non-minimal coupling is absent.

A somewhat similar scheme was introduced recently in \cite{reuterbecker} (see especially subsections 4.1.3 and 4.3.3). The main difference is that in that paper the full physical description of the system is assumed to be contained in the gravitational effective action, with no consideration of the quantum fluctuations below the cutoff scale. The same assumption is made in \cite{litimfalls}, where an explicit identification of the cutoff scale with the size of the black hole is proposed. We take an alternative interpretation of the renormalization group, in which the cutoff scale is an arbitrary parameter separating short and long wavelength modes, and use it as a tool for probing the entanglement interpretation of the long wavelength contribution to the entropy.

The structure of the paper is as follows: In Section 2 we will present the definition of black hole entropy from a canonical ensemble and its connection with the effective action for the gravitational field. In Section 3 we focus on free quantum fields and we introduce the definition of the running effective action dependent on a RG cutoff scale, and 
exhibit 
the renormalization properties of black hole entropy, including a detailed computation for 
the massless scalar field. In Section 4 we present a scheme for extending this idea to 
interacting fields. In Section 5 we discuss a number of issues that impede a direct interpretation of the
IR contributions to the entropy as entanglement entropy.
Section 6 includes a summary of the results and a discussion.

Throughout this paper we work in four dimensions and use units with $\hbar = c = 1$.

\section{Canonical ensemble and entanglement entropy}

This paper will focus on the properties of black hole entropy in a thermal state, 
as defined  by a canonical partition function at fixed temperature:
\begin{equation}\label{Z}
Z=\mathrm{Tr}\, \,\mathrm{e}^{-\beta \mathcal{H}}\,.
\end{equation}
The canonical Hamiltonian $\mathcal{H}$ in (\ref{Z}) includes terms both for the gravitational field and the matter field.
Because of diffeomorphism invariance, $\mathcal{H}$ is a boundary term, when acting on physical states that
satisfy the diffeomorphism constraints. The thermal ensemble is thus specified  
by boundary conditions.
The entropy is computed, as is standard in statistical mechanics, by application of the operator $(\beta\partial_\beta-1)$ to $-\ln Z$. 
This
expression for $Z$ is formally equivalent \cite{gibbonshawk} to the Euclidean path integral 
\begin{equation}\label{fullZ}
Z(\beta,g_B)=\int_{\beta,g_B}\mathcal{D}g\,\mathcal{D}\varphi\,\mathrm{e}^{-S_b[g]-S[g,\varphi]}\,,
\end{equation}
where $S_b$ is a bare action for the gravitational field and $S(g,\varphi)$ is an action for the matter field 
$\varphi$ on background $g$.\footnote{A demonstration of this formal equivalence taking the constraints
into account is given in Ref.~\cite{gross}.} 
The notation $\int_{\beta,g_B}$ represents integration over Euclidean fields 
with the metric $g_B$ fixed and stationary with periodicity $\beta$ in Euclidean time at an outer boundary.
For the matter fields, we are free to choose any boundary condition (e.g. Dirichlet or Neumann) as long as it is 
stationary and compatible with the $\beta$-periodicity. These alternative choices may represent genuinely 
different physical ensembles leading to different results for the entropy. In this paper we will make the simplest 
assumption of Dirichlet boundary conditions, leaving a broader discussion of the issue for  
later work. 
We take the boundary to have a finite size, small enough for the canonical ensemble to be stable \cite{york}.
(Alternatively we could work in asymptotically Anti-de Sitter spacetime\cite{HawkingPage}.)

We can formally integrate out the matter field in (\ref{fullZ}), defining
 \begin{equation}\label{GammaWdefs}
 W[g]=-\ln \int_{\beta, g_B}\mathcal{D}\varphi\,\mathrm{e}^{-S[g,\varphi]}\,\,,\quad\quad \Gamma[g]=S_b[g]+W[g]\,,
 \end{equation}
  so that
\begin{equation}
Z(\beta,g_B)=\int_{\beta,g_B}\mathcal{D}g\,\,\mathrm{e}^{-\Gamma[g]}\,.
\end{equation}
The matter contribution $W$ to the gravitational effective action $\Gamma$ is generally UV divergent. We assume that a regularization scheme for it is in place, and that the bare gravitational couplings (parameters of $S_b$) are adjusted so that the renormalized couplings in $\Gamma$ are finite.

We use now a zeroth-order approximation for quantum gravity, in which the gravitational path integral is evaluated at the saddle point. (This amounts to disregarding from the calculation all graviton fluctuations; in a more full treatment, which we omit for simplicity, they could be included perturbatively among the matter fields $\varphi$.) Therefore we write:
\begin{equation}\label{ZGamma}
Z(\beta,g_B)=\mathrm{e}^{-\Gamma[\bar{g}(\beta,g_B)]}\,.
\end{equation}
Here $\bar{g}(\beta,g_B)$ is the metric that solves the equations of motion derived from $\Gamma[g]$, with the boundary conditions that we have set. We assume that the renormalized cosmological constant is zero or 
negligible,
and that the higher-order in curvature terms of $\Gamma$ (which include nonlocal terms) can likewise be neglected in the regime of interest. Then $\Gamma[g]$ is comprised only of the bulk curvature term and the corresponding Gibbons-Hawking boundary term\footnote{It has recently been pointed out \cite{reuterbecker} that the $G_{\mathrm{ren}}$ that appears in the boundary term of the effective action might not be the same as the one in the bulk term when the matter field is non-minimally coupled to the curvature. For the moment, we assume that this is not the case and that bulk and boundary $G_{\mathrm{ren}}$ are equal, as is the case, for example, when $\varphi$ is the minimally coupled scalar field.}:
\begin{equation}
\Gamma[g]=-\frac{1}{16 \pi G_{\mathrm{ren}}}\int\sqrt{g}R\,-\frac{1}{8 \pi G_{\mathrm{ren}}}\int_{\partial M}\sqrt{h}K \,.
\end{equation}

 Now we specialize the discussion to the ensemble defined by a spherically symmetric boundary metric $g_B$, which is a 2-sphere of radius $r_B$. We also work in four dimensions, though most of our results generalize straightforwardly to $d$ dimensions.
The on-shell metric $\bar{g}$ is Euclidean Schwarzschild:
 \begin{equation}\label{Sch}
\overline{\mathrm{d}s^2}=\left(1-\frac{\overline r_+}{r}\right)\mathrm{d}t^2
+\left(1-\frac{\overline r_+}{r}\right)^{-1}\mathrm{d}r^2+r^2\,\mathrm{d}\Omega^2\,,
\end{equation}
with the horizon 
radius 
$\overline r_+(\beta,r_B)$  defined
so that (i) there is no conical singularity at $r=\overline r_+$ and (ii) the ensemble is
stable ($dr_+/d\beta<0$). These conditions imply \cite {york} 
that $\overline r_+$ is the larger root of
the equation:
\begin{equation}\label{relbetar}
\beta=4\pi\,r_+\left(1-\frac{r_+}{r_B}\right)^{1/2}.
\end{equation}
The ``box'' at the boundary $r_B$ stabilizes the ensemble by giving the black hole a 
positive heat capacity.\footnote{If the black hole grows a 
little bit, its horizon is closer to the box, so less redshifting  of temperature 
occurs from the horizon to the box. If $r_B<3r_+/2$, then this suppression of 
redshifting makes the temperature of the larger black hole 
higher than $1/\beta$  at the box.}

When evaluated on this metric, the bulk term vanishes and the boundary term  yields: 
 \begin{equation}\label{onshell}
 \Gamma[\bar{g}]=\frac{1}{ G_{\mathrm{ren}}}(3\pi\, \overline r^2_+-4\pi\,r_B\, \overline r_+),
 \end{equation}
The entropy is given by the standard thermodynamical formula
\begin{equation}\label{onshellformula}
S_{BH}=-\left(\beta\partial_\beta-1\right)\ln Z(\beta,r_B)=\left(\beta\partial_\beta-1\right)\Gamma[\bar{g}(\beta,r_B)]\,.
\end{equation}
When applied to (\ref{onshell}), this results in the renormalized Bekenstein-Hawking formula:
\begin{equation}
S_{BH}=\frac{A_{}}{4G_{\mathrm{ren}}}\,,
\end{equation}
where $A_{}=4\pi \bar r_+^2$ is the horizon area.

The above 
approach to evaluating the matter field contribution to $S_{BH}$ has been dubbed in the past the ``on-shell'' computation or the ``thermodynamical'' computation \cite{reg1,reg2}. 
 Within this procedure, the fact that the full entropy including quantum corrections is expressed by the Bekenstein-Hawking formula involving the renormalized Newton constant $G_{\mathrm{ren}}$ is an immediate consequence of the renormalization  of the effective action, as was emphasized in \cite{larsen}. Note that this also implies that the so-called ``species problem'' (the dependence of the quantum contribution to black hole entropy on the kinds of existing quantum fields, apparently contradicting the universality of the Bekenstein-Hawking formula \cite{sorkin1}) is moot because in terms of the renormalized value $G_{\mathrm{ren}}$, this formula is always correct regardless of the field species (which affect the relation of $G_{\mathrm{ren}}$ to the unobservable bare value). 
 
In contrast, when computing the contribution of the matter fields as entanglement entropy on the black hole background
\cite{sorkin1,bkls,sred,frolov93}, this renormalization property is much less apparent. We shall therefore discuss in more detail 
the relation to entanglement entropy computations.

To compute the entanglement entropy contribution to black hole entropy, we consider a 
minimally coupled 
quantum field $\varphi$ on a Schwarzschild background. The entanglement entropy across the event horizon is defined by 
\begin{equation}
S_{\mathrm{ent}}=-\mathrm{Tr}\, \rho_{\mathrm{out}} \ln \rho_{\mathrm{out}}\,,
\end{equation}
where $\rho_{\mathrm{out}}$ is the restricted density matrix for the external region, with the internal states traced over. Using a Euclidean path integral representation for $\rho$, this can be rewritten \cite{suss2, wilzec, solod2}  as the operation $(\alpha \partial_\alpha-1)$ applied to the matter contribution $W_\alpha$ to the effective action, computed on the Euclidean Schwarzschild background with a conical singularity introduced at the horizon.  Here $2\pi\alpha$ is the periodicity of the angular ``time'' coordinate and $\alpha$ is set to 1 after the differentiation:
\begin{equation}
S_{\mathrm{cone}}=\left(\alpha \partial_\alpha-1\right)W_\alpha\Big|_{\alpha=1}\,.
\end{equation}
Actually this ``conical'' procedure yields the entanglement entropy only for minimally coupled fields. For non-minimally coupled fields, $S_{\mathrm{cone}}$ 
is not equal to $S_{\mathrm{ent}}$,  as it includes an extra, contribution 
interpretable \cite{waldnonmin,donn} as 
the expectation value of a term in the 
Wald entropy \cite{wald}. 
Our subsequent discussion in this section 
assumes minimal coupling.  We shall return to expand a bit on this issue in Section 5. 

The use of a background with a deficit angle $2\pi(1-\alpha)$ is equivalent to using  periodic quantum fields whose physical periodicity $\beta$ at a radius $r_B$ is related to the horizon parameter $r_+$ by
\begin{equation}\label{rbar}
\beta=4\pi\,r_+\,\alpha\left(1-\frac{r_+}{r_B}\right)^{1/2}\,,
\end{equation}
rather than by (\ref{relbetar}). Hence $\alpha$ and $\beta$ are proportional at fixed $r_+$, and the entanglement entropy is equivalently given by
\begin{equation}\label{offshell}
S_{\mathrm{ent}}=\left(\beta \partial_\beta-1\right)W[g(\beta,r_B,r_+)]\Big|_{r_+=\overline r_+}\,.
\end{equation}

Comparing with (\ref{onshellformula}) and recalling (\ref{GammaWdefs}), the relationship between the entanglement black hole entropy and the thermodynamical black hole entropy becomes clear: The entanglement entropy (when computed using as background the metric $\bar{g}$) differs from the matter contribution to the thermodynamical entropy, only in that the variation with respect to $\beta$ is done keeping $r_+$ fixed (thus involving the introduction of a deficit angle), instead of taking into account the dependence of $r_+$ on $\beta$ 
in 
the on-shell solution
Hence the dubbing of this as the ``off-shell'' method in \cite{reg1}.

However, it can be argued that the two methods give actually the same results. First, consider again (\ref{onshellformula}), and recast the dependence of $\Gamma[\bar{g}(\beta,r_B)]$ as $\Gamma[g(\beta,r_B,\bar{r}_+(\beta,r_B))]$. Then the ``total'' $\beta$-derivative appearing in  (\ref{onshellformula}) can be unpacked as 
\begin{equation}\label{totalder}
\partial_\beta\Gamma[\bar{g}(\beta,r_B)]=\Bigl(\partial_\beta \Gamma[g(\beta,r_B,r_+)]
+\partial_{r_+}\Gamma[g(\beta,r_B,r_+)]\partial_\beta{r}_+\Bigr)\Big|_{r_+=\bar{r}_+}\,.
\end{equation}
The last term should vanish, because $\partial_{r_+}\Gamma[g(\beta,r_B,r_+]$ expresses a variation of the action with respect to the metric with fixed boundary conditions, which is zero at the on-shell value of $r_+$ (a stationary point of $\Gamma$)\footnote{This argument has been made previously in \cite{solod3}; see also \cite{frolova4}.}. Indeed, this can be checked with a direct computation of this derivative evaluated on the deficit-angle version of Schwarzschild, where the conical singularity is accounted for as an extra term in the curvature scalar:
\begin{equation}
R=\bar{R}+4\pi(1-\alpha)\delta_\Sigma\,.
\end{equation}
So we conclude that the thermodynamical entropy can be expressed using just the first term of (\ref{totalder}), which includes the same kind of ``off-shell'' partial derivative as (\ref{offshell}). Hence we have:
\begin{align}
S_{BH}&=\left(\beta \partial_\beta-1\right)\Big(S_b[g(\beta,r_B,r_+)]+W[g(\beta,r_B,r_+)]\Big)\Big|_{r_+=\bar{r}}
\nonumber\\
&=\left(\beta \partial_\beta-1\right)S_b[g(\beta,r_B,r_+)]\Big|_{r_+=\bar{r}_+}\,
+S_{\mathrm{ent}}\,.
\end{align}
Therefore the entanglement entropy must share the renormalization property of the thermodynamical entropy: its divergences should be absorbable in a redefinition of the bare gravitational couplings, as it occurs for $W$ in the thermodynamical entropy.

Note that the equality requires the entanglement entropy to be computed using as background spacetime the metric $\bar{g}$ that solves the quantum-corrected equations of motion.  In practice, this metric is assumed to be Schwarzschild (or another known black hole solution to the gravitational theory) expressed in terms of the observable, low-energy couplings.

An issue we have mentioned but not dedicated proper attention to is the need to regularize $W$ (the matter contribution to $\Gamma$) to make it finite. Insofar as $W$ is divergent, the whole argument is not rigorously defined. It would clearly be preferable if the renormalization properties of the entropy 
and the relation to entanglement entropy
could be studied by manipulation of manifestly finite quantities only. In the rest of this paper, we achieve this by introducing a Wilsonian renormalization group scale $k$, at which the 
entropy can be decomposed into 
respective contributions 
from the gravitational action and the matter action. 
The flow of these contributions as $k$ changes is then well-defined. 
In the next section we study this flow for free matter fields, and in section 4 matter
interactions are included. In section 5 the viability of interpreting the matter 
contributions as entanglement entropy is probed.

\section{RG flow of black hole entropy for free fields}

In this section we will show how the black hole entropy, in the ``thermodynamical'' framework presented above, can be described in a way that makes its renormalization properties clear avoiding the handling of divergences. In the spirit of the Wilsonian interpretation of the renormalization group \cite{wilson}, the idea is to introduce a cutoff scale $k$ and to integrate out only the quantum modes above that scale. 

Let us start by going back to (\ref{ZGamma}), after the zeroth-order approximation for the gravitational path integral has been made, and replace $\Gamma[\bar{g}]$ by its definition:
\begin{align}\label{ZSW}
Z(\beta,r_B)&=\mathrm{e}^{-S_b[\bar{g}(\beta,r_B)]-W[\bar{g}(\beta,r_B)]}\nonumber\\
&=\mathrm{e}^{-S_b[\bar{g}(\beta,r_B)]}\int_{\beta,r_B}\mathcal{D}\varphi\,\mathrm{e}^{-S[\bar{g}(\beta,r_B),\varphi]}\,.
\end{align}
The path integral should be assumed to contain an implicit covariant UV regulator, for example a short-distance cutoff $\epsilon$ in the heat kernel expansion. If $\varphi$ is a free field, assumed for illustration to be a scalar, then its action takes the general form\footnote{We assume that there is no boundary term in the matter action, or more exactly that it vanishes when the boundary conditions are imposed.}
\begin{equation}
S[g,\varphi]=-\frac{1}{2}\int\sqrt{g}\,\varphi \Delta_g\varphi\,,
\end{equation}
where $\Delta_g$ is an elliptic operator depending on the metric $g$; for example $\Delta=-\nabla_g^2$ for a
massless
minimally coupled scalar field, and $\Delta=-\nabla_g^2+\xi R(g)$ for a nonminimally coupled one. Here $\nabla_g^2$ is the Laplacian operator on background $g$. The matter contribution $W$ to the effective action is given by the one-loop determinant:
\begin{equation}\label{oneloop}
\Gamma[g]=S_b[g]+\frac{1}{2}\mathrm{Tr}_\epsilon \ln \Delta_g\,,
\end{equation} 
which can be computed with standard heat kernel expansion techniques. With the regulator $\epsilon$ removed, the right hand side is divergent; leaving it in place, we can make the couplings of the bare action $S_b$ dependent on $\epsilon$ in such a way that the effective action $\Gamma$ and the physics derived from it 
are 
independent of $\epsilon$.

We now want to introduce an intermediate RG scale $k< 1/\epsilon$. One simple way of doing this is using an additive cutoff as introduced by Wetterich \cite{wette} to study the exact renormalization group. The gravitational effective action at scale $k$ is defined in general as:
\begin{equation}\label{defGammak}
\Gamma_k[g]=S_b[g]-\ln\int \mathcal{D}\varphi\,\mathrm{e}^{-S[g,\varphi]-\frac{1}{2}\int\sqrt{g}\,\varphi\left[\mathcal{R}_k(\Delta_g)\right]\varphi}\,.
\end{equation}
Here $\mathcal{R}_k(\Delta)=k^2 r(\frac{\Delta}{k^2})$, with $r(z)$ being a function that satisfies the properties $r(0)=1$ and $r(z)=0$ for $z\gtrsim 1$. This implies that the $\mathcal{R}_k$ term serves as an IR cutoff in the path integral, suppressing from $\Gamma_k$ the contribution of the modes with eigenvalue $p^2<k^2$. Hence $\Gamma_k$ is an ``average'' effective action that only incorporates the effect on the gravitational couplings of fluctuations on length scales smaller than $k^{-1}$. For $k\rightarrow 0$, $\Gamma_k$ approaches the full effective action 
$\Gamma$.\footnote{Note 
that with the cutoff function $\mathcal{R}_k$ the suppression of the modes with $p^2<k^2$ is
not complete; it is rather like giving them a mass $\sim k^2$.}

In 
the free field case, $\Gamma_k$ can be computed exactly as a modified one-loop determinant:
\begin{equation}\label{oneloopk}
\Gamma_k[g]=S_b[g]+\frac{1}{2}\mathrm{Tr}_\epsilon \ln \left[\Delta_g+\mathcal{R}_k\left(\Delta_g\right)\right]\,,
\end{equation}
which by differentiation with respect to RG flow ``time'' $t=\ln k$ yields the well-known RG flow equation
 \cite{wette}
\begin{equation}\label{ERGE}
\partial_t \Gamma_k = \frac{1}{2}\mathrm{Tr}\left[\frac{\partial_t\mathcal{R}_k}{\Delta_g+\mathcal{R}_k\left(\Delta_g\right)}\right]\,.
\end{equation}
Comparing (\ref{oneloopk}) and (\ref{oneloop}), we get the following expression for the difference between the total effective action and the effective action at scale $k$:
\begin{equation}\label{actions} 
\Gamma[g]-\Gamma_k[g]=\frac{1}{2}\mathrm{Tr}\ln\left[\frac{\Delta_g}{\Delta_g+\mathcal{R}_k\left(\Delta_g\right)}\right]\equiv W_k[g]\,.
\end{equation}
Notice that we can drop the UV cutoff $\epsilon$ in this expression, because the trace has now acquired a lower intrinsic UV cutoff at scale $k$ since $\mathcal{R}_k=0$ for eigenvalues $p^2\gtrsim k^2$; the right hand side of (\ref{actions}) is well-defined regardless of any overall UV cutoff for the theory. Evaluating at the on-shell Euclidean Schwarzschild metric $\bar{g}=\bar{g}(\beta,r_B)$ and applying the operator $(\beta\partial_\beta-1)$ we obtain the relation for the entropies:
\begin{equation}\label{entropies} 
S_{BH}-S_{BH}^{(k)}=\Big(\beta\partial_\beta-1\Big)W_k\,.
\end{equation}
$S_{BH}$ is the total black hole entropy, incorporating the bare gravitational contribution and the total
effect of the quantum matter fields. $S_{BH}^{(k)}$ is the ``effective gravitational entropy at scale $k$'',  derived from the effective action $\Gamma_k[g]$; it incorporates both the bare gravitational contribution and the effect that the high-frequency quantum modes have on the renormalization of the gravitational couplings. The right hand side encompasses the effect of the low-frequency modes, whose contributions had been suppressed from $\Gamma_k$. The total entropy $S_{BH}$ is independent of the sliding scale $k$, which partitions it into two complementary contributions. Thus the renormalization properties of the entropy are made clear without need to worry about the global UV cutoff and the treatment of divergences. $S_{BH}$ is expressed in terms of low-energy, observable couplings, and the other terms in the equation differ from it by a finite RG scale $k$, which in principle can receive a physical interpretation. 

To investigate whether the right hand side of (\ref{entropies}) can be interpreted as the entanglement entropy of the modes below scale $k$, let us evaluate it in a concrete example. We consider a minimally coupled 
massless
scalar field $\varphi$, so $\Delta_g=-\nabla^2_g$. We choose as cutoff function the ``optimized cutoff'' introduced by Litim \cite{litim}, which is given by:
\begin{equation}
\mathcal{R}_k(\Delta)=k^2\,r\left(\frac{\Delta}{k^2}\right);\quad\quad r(z)=(1-z)\Theta(1-z)\,.
\end{equation}
We have in this case that the operator traced upon in $W_k$ 
(\ref{actions}) 
is exactly zero for eigenvalues above $k^2$:
\begin{equation} 
W_k[\bar{g}]=\frac{1}{2}\mathrm{Tr}\left\{\ln\left[\frac{-\nabla^2_{\bar{g}}}{k^2}\right]\Theta\left[k^2-(-\nabla^2_{\bar{g}})\right]\right\}\,.
\end{equation}

We can use the procedure described in the Appendix A of \cite{codello}, adapted for a 4-dimensional manifold with boundaries, for using the heat kernel expansion to compute the trace of an arbitrary function of our operator, $F(\Delta)$.
Assuming Dirichlet or Neumann boundary conditions are imposed, one has 
\begin{equation}
\mathrm{Tr}[F(\Delta)] = \frac{1}{16\pi^2}\left[a_0(\Delta)Q_2(F)+a_1(\Delta)Q_{\frac{3}{2}}(F)+a_2(\Delta)Q_1(F)+ \cdots \right]\,,
\end{equation}
where the coefficients $Q_n(F)$ are defined as
\begin{equation}
Q_n(F)=\frac{1}{\Gamma(n)}\int \mathrm{d}z\, z^{n-1}\,F(z)\,,
\end{equation}
and the heat kernel coefficients $a_n$ take the form \cite{HK}:
\begin{eqnarray}
a_0(\Delta)&=&\int\sqrt{g}\\
a_1(\Delta)&=&\sqrt{\frac{\pi}{4}}\int_{\partial M}\sqrt{h}\\
a_2(\Delta)&=&\frac{1}{6}\left[\int\sqrt{g} R +2\int_{\partial M}\sqrt{h}\,K\right].
\end{eqnarray}
For $F(z)=\ln\left[\frac{z}{k^2}\right]\Theta\left[k^2-z\right]$, 
we have $Q_n=-k^{2n}/n^2\Gamma(n)$. 
Hence we obtain:
\begin{equation}\label{Wcoeffs} 
W_k[g]=-\frac{1}{32\pi^2}\left\{\int\sqrt{g}\left[\frac{k^4}{4}+\frac{k^2}{6} R\right]
+\int_{\partial M}\sqrt{h}\left[\frac{4 k^3}{9}+\frac{k^2}{3}K\right]+\cdots\right\}\,.
\end{equation}

The next step is to evaluate at the on-shell Euclidean Schwarzschild metric $\bar{g}(\beta,r_B)$ and to compute the right hand side of (\ref{entropies}). The bulk curvature term is zero, and the bulk and boundary volume terms are proportional to $\beta$ and hence vanish upon application of the $(\beta\partial_\beta-1)$ operator. The boundary $K$ term gives a result proportional to the event horizon area. Hence, neglecting higher-order curvature terms, (\ref{entropies}) takes the form
\begin{equation}\label{freephiren}
\frac{A_{}}{4G_0}-\frac{A_{}}{4G_k}=\frac{A_{}}{4}\left(\frac{k^2}{12\pi}\right)\,.
\end{equation}
This equation 
can be read in two ways. 
On one hand, canceling the $A_{}/4$'s, it just expresses the RG running of $G$ due to the quantum corrections induced by the scalar field,
which was already implicit in (\ref{Wcoeffs}). 
On the other hand, we can interpret it as expressing two contributions to black hole entropy: For any scale $k$, the total black hole entropy $A_{}/4G_0$ (where $G_0$ is the fully renormalized Newton constant) can be partitioned in two contributions, the effective gravitational entropy at scale $k$ (which is $A_{}/4G_k$) and the 
contribution 
of the scalar field's modes that are below $k$, which is $A_{} \,k^2/48\pi$. 
When sliding the RG scale $k$, the balance of the entropy is shifted between the two terms, leaving the total entropy unchanged.

It would seem natural to regard the contribution of the lower modes as corresponding to 
their horizon entanglement entropy. It has the same form as the the total entanglement
entropy calculated with a UV momentum cutoff $\Lambda$, with the intermediate scale 
$k$ playing the role of $\Lambda$.  Of course the 
precise expression for the running of $G_k$ depends on the cutoff function $\mathcal{R}_k$. A different choice of regulator would lead to a different numerical coefficient of $A_{} \,k^2$. 
In itself this does not seem problematic for the entanglement entropy interpretation,
since it just reflects the implementation of the partitioning of the contributions
from degrees of freedom above and below the RG scale.

In section 5 we shall discuss some other questions concerning
the justification of the entanglement interpretation of the contribution
of the lower modes. First, however, we consider how the preceding
analysis must be modified in order to account for interactions
of the matter degrees of freedom.

\section{Interacting fields}

In this section we will discuss how our framework can be extended to interacting quantum fields. It turns out that the distinction at a scale $k$ between the gravitational entropy and the 
contribution
from the modes below $k$ can be defined as we did for free fields, though it is much more difficult to write down the exact form of each contribution for a given example. There are also 
further
interpretational issues, which will be addressed in Section 5.

Let $S[g,\varphi]$ be a bare action for the quantum field $\varphi$ on the gravitational background $g$. Equation (\ref{ZSW}) for the black hole partition function is true regardless of whether $S$ contains interactions. We can, as before, introduce an additive cutoff $\mathcal{R}_k$ and define the running gravitational effective action $\Gamma_k[g]$ by (\ref{defGammak}), and define the gravitational part of the entropy at scale $k$ by applying the $(\beta\partial_\beta -1)$ operator to $\Gamma_k[\bar{g}]$. However, since for interacting $\varphi$ the one-loop determinant is not an accurate evaluation of the effective action, we are lacking a compact expression like (\ref{entropies}) for the contribution of the modes below $k$. 
Integrating out the upper modes not only produces running in the gravitational effective action, but also 
the Wilsonian effective action for $\varphi$ depends on the scale $k$. In the following we elaborate on how to quantify this running and obtain expressions for interacting fields as close as possible in spirit to (\ref{entropies}).

Let us start, again, with the total partition function, with an overall UV cutoff implicitly in place 
with a short-distance regulator $\epsilon$. We isolate the kinetic term in the 
matter
action and define as $S_b[g,\varphi]$ the sum of the bare gravitational action $S_b[g]$ and the non-kinetic terms of the bare matter action.
Then the partition function is given by 
\begin{equation}\label{Zint}
Z[g]=\mathrm{e}^{-\Gamma[g]}=\int\mathcal{D}\varphi\,\mathrm{e}^{-\frac{1}{2}\int\varphi (-\nabla_g^2) \varphi\,-\,S_b[g,\varphi]}\,.
\end{equation}
We now introduce the IR cutoff function $\mathcal{R}_k$, with the same properties as in 
the previous
section,  
defining the Wilsonian effective action 
at scale $k$
by:
\begin{equation}\label{wilsoniandef}
\mathrm{e}^{-S_k[g,\phi]} =\int\mathcal{D}\varphi\,\mathrm{e}^{-\frac{1}{2}\int \varphi \left(-\nabla_g^2+\mathcal{R}_k(-\nabla_g^2)\right)\varphi- S_b[g,\phi+\varphi]}\,.
\end{equation}
The purely gravitational part of $S_k$ 
coincides with our previous definition of $\Gamma_k$, and we 
shall be decomposing the entropy into the contribution from $\Gamma_k$ and that from the remainder, 
which captures the physics of the lower modes of the matter field $\phi$ on the background $g$. 
To that end, we introduce the notation
\begin{eqnarray}\label{defStilde}
\Gamma_k[g]&:=&S_k[g,\phi=0]\\
\tilde{S}_k[g,\phi] &:=& S_k[g,\phi]-\Gamma_k[g]\,.
\end{eqnarray}

The action $S_k$ includes the effects of the modes ``above $k$", so the partition function
(\ref{Zint}) should be expressible as a path integral over the modes ``below $k$" using 
this action. We find such an expression by assuming it can be written in the form
\begin{equation}\label{lowerZ}
Z[g]=N_k[g]\int\mathcal{D}\phi\,\mathrm{e}^{-\frac{1}{2}\int\phi P^{-1}_k \phi\,-\,S_k[g,\phi]}\,,
\end{equation}
for some suitable choice of normalization $N_k$ and IR propagator $P_k$. 
Substituting (\ref{wilsoniandef}) into (\ref{lowerZ}), shifting one of the field variables so that its integral is Gaussian, and performing the integral, 
we find that (\ref{Zint}) is recovered with the following definitions of $N_k$ and $P_k$:
\begin{equation}\label{defN}
N_k[g]=\mathrm{det}^{1/2}\left[\frac{\left(-\nabla_g^2+\mathcal{R}_k(\nabla_g^2)\right)^2}{\mathcal{R}_k(\nabla_g^2)}\right]\,,
\end{equation}
\begin{equation}\label{defP}
P_k(\nabla_g^2)=\left(\frac{\mathcal{R}_k(\nabla_g^2)}{-\nabla_g^2}\right)\frac{1}{-\nabla_g^2+\mathcal{R}_k(\nabla_g^2)}\,.
\end{equation}
Note that since $\mathcal{R}_k$ vanishes for modes above $k$, so does the propagator $P_k$ and hence the path integral in (\ref{lowerZ}) acquires a UV cutoff at this scale. We stress that (\ref{lowerZ}) is 
identical to the full path integral, with the information about the upper modes encoded in the Wilsonian effective action $S_k[g,\phi]$, which is obtained from integrating them out according to (\ref{wilsoniandef}).

The definition of $S_k$ given by (\ref{wilsoniandef}) is purely formal however, and unsuitable 
for analyzing the entropy of the black hole at scale $k$. In first place, the path integral cannot be computed in a closed form for an interacting theory. Moreover, the expression in terms of a bare action and a path integral which is divergent requires an explicit regularization procedure to deal with divergences. This goes against the spirit of our approach, based on analyzing the difference between the expressions for the entropy obtained at different effective scales, in terms of finite, effective quantities only.

The right tool for these purposes is an RG flow equation, detailing how $S_k[g,\phi]$ changes with the scale $k$ in a local way. This is the Polchinski equation \cite{polch}, which 
in the present case
takes the form: 
\begin{eqnarray}
\dot{S}_k&=&\dot{\Gamma}_k+\dot{\tilde S}_k\\
&=&\frac{1}{2}\left\{\frac{\delta \tilde S_k}{\delta\phi}\cdot \dot{P}_k\cdot\frac{\delta \tilde S_k}{\delta \phi}-\mathrm{Tr}\left[\dot{P}_k\cdot\frac{\delta^2\tilde S_k}{\delta\phi\delta\phi}\right]+\mathrm{Tr}\left[\dot{P}_k(-\nabla_g^2+\mathcal{R}_k)\right]\!\right\}\!\!,
\label{Polch1}
\end{eqnarray}
where the overdots represent $k$-derivatives, and the center dot notation ($\cdot$) is explained in Appendix A. Since this equation differs by the last term from the flat space form that is derived in standard presentations of the Wilsonian renormalization group \cite{morris, polonyi}, we detail in Appendix A how it is obtained using (\ref{wilsoniandef})  as starting point.

Expanding the flow equation with a systematic approximation method (e.g. a derivative expansion) would give beta functions for each of the gravitational couplings in $\Gamma_k$ and of the field couplings in $\tilde{S}_k$. Let us assume we are in possession of a solution to these flow equations for the couplings (found, perhaps, with numerical techniques, and using as initial condition for the flow a known form of the effective action at low energies). This 
would then allow us to write down the form of $\Gamma_k[g]$ and $\tilde{S}_k[g,\phi]$ for any given value of the RG scale $k$. 
The log of the partition function (\ref{lowerZ}) is then expressible in terms of these 
quantities as 
\begin{equation}\label{ZGammaWint}
-\ln Z[g] =\Gamma[g]= \Gamma_k[g] +W_k[g],
\end{equation}
with 
\begin{equation}\label{Wint}
W_k[g] = -\ln \left[N_k[{g}]\int\mathcal{D}\phi\,\mathrm{e}^{-\frac{1}{2}\int\phi P^{-1}_k \phi\,-\,\tilde{S}_k[{g},\phi]}\right]\,.
\end{equation}
In this way the free energy is decomposed into finite, 
purely gravitational and matter parts in a scale-dependent manner.
This decomposition generalizes the one (\ref{actions}) discussed previously for free fields.

The saddle point approximation to the total entropy of the thermal ensemble
is obtained using the partition function evaluated at the metric $\bar{g}(\beta,r_B)$ given by (\ref{Sch}),
which is a solution to the full effective action at $k=0$, i.e.\ with all fluctuations integrated out. 
Using the decomposition (\ref{ZGammaWint}, \ref{Wint}) for $Z$, we obtain:
\begin{align}\label{entropiesint}
S_{BH}&= (\beta\partial_\beta-1) (\Gamma[\bar{g}]+W_k[\bar{g}])\nonumber\\
&=: S_{BH}^{(k)}\,+(\beta\partial_\beta-1) W_k[\bar{g}]\,.
\end{align}
The term
$S_{BH}^{(k)}$, is (as before) what we define as the gravitational black hole entropy at scale $k$. The second term, 
with the definition (\ref{Wint}),
generalizes the right hand side of (\ref{entropies}) 
with its definition (\ref{actions}),
 and encompasses the contribution to the entropy of the quantum modes below scale $k$, as computed with the appropriate Wilsonian action for them. It can easily be checked that for the case of free fields, where the path integral is Gaussian and can be done exactly,
  (\ref{Wint}) reduces precisely to (\ref{actions}) so that both expressions for the entropy agree. 

Once more, the renormalization properties of the entropy are made explicit without the need of specifying the global UV cutoff and handling divergences; a global implicit regulator $\epsilon$ is needed to make (\ref{Zint}) well-defined, but it drops from the calculations and when we reach (\ref{entropiesint}) we are dealing only with finite physical quantities. The total entropy, the right hand side of (\ref{entropiesint}), does not depend on the scale $k$, but when the scale is shifted the relative balance of the two contributions on the right hand side is changed, as the gravitational and non-gravitational part of the effective action flow according to (\ref{Polch1}).

\section{Interpretation of the IR contribution to the entropy}

We have studied in the previous two 
sections
the RG flow of black hole
entropy contributions coming from above and below a running scale $k$.
We obtained expressions
(\ref{entropies}) for free fields and (\ref{entropiesint}) for interacting
fields, with the associated definitions (\ref{actions}, \ref{Wint}) for $W_k$ in each case. 
These decompositions are interesting in their own right, 
as they illustrate how a black hole entropy computation tracks the 
RG flow of gravitational couplings. But our main motivation for undertaking
this exercise was to test the notion that black hole entropy is, at least 
in part, entanglement entropy of quantum fields, in a controlled setting 
where no divergent quantities arise and where properties of the UV 
completion of the theory are irrelevant. In this framework,
the results may have a more direct and less ambiguous physical 
interpretation.

In particular, the tempting interpretation of (\ref{entropies}) and (\ref{entropiesint}) (as discussed briefly 
with regard to (\ref{freephiren})), is that the IR contribution to the entropy, 
\begin{equation}\label{IRcontribution}
(\beta\partial_\beta-1) W_k[\bar{g}]\,,
\end{equation} 
can
be identified with the entanglement entropy of the modes below scale $k$.
There are a number of considerations that complicate this interpretation however. 
We will now discuss them.

 \subsection{Contact terms}

As mentioned in section 2, for non-minimally coupled scalar fields, and 
perhaps for gauge fields and gravitons, the ``conical entropy"
(i.e.\ the entropy of
the thermal partition function defined on a space that acquires
a conical deficit when the Euclidean period is varied off-shell) contains a
contribution from the tip of the cone, the so-called contact term, that 
does not appear to admit a statistical interpretation.\footnote{\label{0modes}However, 
in Refs.~\cite{waldnonmin, Fursaev:1998hr,Fursaev:1999jq} 
a statistical interpretation was proposed in terms of zero energy modes
localized at the horizon.} 
This contact term arises also in the contribution from the modes below scale $k$, so 
in general that contribution would not consist only of entanglement entropy. 
In a specific interacting model in 1+1 dimensions \cite{onmodel}, 
it was illustrated how 
non-minimal coupling and the associated contact term can arise from 
a Wilsonian effective action when some degrees of freedom above
a certain mass scale are integrated out. 
Hence it is possible, in a given setting, that the contact term is a stand-in for 
an entanglement contribution, but that need not be so.

The contact term is a hybrid between a 
gravitational and a quantum contribution. 
It can be interpreted 
\cite{waldnonmin,donn} 
as a term in the Noether charge, i.e.\
in the Wald entropy \cite{wald}, involving the expectation value of the 
squared matter field with the cutoff $k$.
Thus, in the presence of non-minimal coupling, we should
refine our conjecture about the decomposition of the total entropy at scale $k$
into gravitational and entanglement contributions. The gravitational
part must include all contributions from the Noether potential at scale $k$. 
For example, for the scalar field with an $R\varphi^2$ coupling in $\tilde{S}_k$, 
the gravitational part would arise both from this term and from $\Gamma_k$.
Since $\varphi$ is a fluctuating quantum field, this makes the Noether potential
an operator rather than a classical quantity. The gravitational part in the 
conjecture involves the expectation value of this operator.

\subsection{Euclidean vs.\ Lorentzian RG scale}

The running scale $k$ in our calculations is defined as a cutoff in the 
eigenvalues of the Euclidean Laplace operator. In the Euclidean domain, 
we have a clean separation between 
two contributions to the entropy: the term of (\ref{entropiesint}) coming from 
$\Gamma_k$ 
represents the ``gravitational entropy" at the scale $k$, and the one involving 
$\tilde{S}_k$ 
represents the contribution of the Euclidean modes below $k$. 
But what exactly do these terms correspond to in the Lorentzian domain, where the entanglement entropy is fundamentally defined? 
There the subsystem of interest would be defined by a cutoff in the eigenvalues of the spatial Laplace operator, in a 3+1 decomposition\footnote{See \cite{elisalor} for a specific implementation.}. In a thermal state with temperature of the order of the cutoff, the two procedures should yield qualitatively similar results. Something like this seems to be the case in the black hole setting: the near horizon part of the ``lower'' contribution to the entropy is dominated by momenta of order $k$ at a distance of order $k^{-1}$ from the horizon, where the fluctuations have a local temperature $\sim k$.
However, the precise relation between the quantity (\ref{IRcontribution}), defined by a Euclidean cutoff, 
and the entanglement entropy of a subset of the Lorentzian quantum fluctuations, remains to be 
fully clarified.

\subsection{Uncertainty relation between horizon location and momentum cutoff}

The notion of ``horizon entanglement entropy" refers to the von Neumann entropy of the 
reduced density matrix of the exterior degrees of freedom. We are here considering
this notion in the presence of a momentum cutoff. If the calculation were
strictly in the Lorentzian domain, the limitation on momenta would presumably
imply that the separation of degrees of freedom on one side of the 
horizon could not be arbitrarily sharp, and in fact would be fuzzy at the 
scale of the inverse momentum cutoff. Since the entanglement entropy
is dominated by the contributions at the shortest scales, this means that there
would be an order unity fuzziness in its value, computed in this fashion.
This dependence on the cutoff would not be so disturbing if it were a 
feature of a UV regulation of an otherwise undefined quantity, but  
one might have hoped for a more
precise definition of the entanglement entropy at the RG scale $k$. 

We do not encounter this fuzziness in our computation. 
The unregulated Euclidean path integral for the partition function can be viewed 
as a formal computation of the trace of $\exp(-\beta H_{ext})$, where 
$H_{ext}$ is the Hamiltonian for the degrees of freedom exterior to  
the horizon. However when this path integral is filtered to include only the modes
with momentum below $k$, it is no longer exactly the trace of an operator on 
the exterior Hilbert space. For this reason, 
what we compute in (\ref{IRcontribution}) is not
precisely the entropy of a reduced density matrix corresponding to a subset of the exterior modes.

\subsection{Momentum entanglement}

For interacting theories, a further complication arises. In the ground
state, degrees of freedom with different
momenta are entangled. In Minkowski spacetime, the reduced 
vacuum density matrix
for IR degrees of freedom 
below a scale $k$
has an entanglement entropy per unit volume.
This was computed perturbatively for various theories in various dimensions
in Ref.~\cite{momentum}, but it is explained there 
that 
for some theories
the perturbative calculation is not adequate. 
Nevertheless, we can estimate that a lower bound for theories in 3+1 dimensions should scale as
$\lambda^2 k^3$, where $\lambda$ is the coupling constant. 
The actual result for a given theory might involve some power of 
a higher energy scale $M$ and logarithms of the ratio $M/k$. 

Momentum entanglement will also play a role for the reduced 
density matrix of IR modes outside a black hole horizon. If we restrict 
attention to the volume of space at a proper distance $l$ from a spherical
horizon, the 
lower bound for 
momentum entanglement entropy would scale as 
$\lambda^2 k^3 l r_+^2$, whereas the horizon entanglement entropy scales as 
$k^2 r_+^2$, hence the former dominates unless $\lambda^2 k l\lesssim 1$. 
If $\lambda$ is much smaller than unity, 
the momentum entanglement contribution could 
potentially 
be suppressed by focusing only on a region of radial width $l<(\lambda^2 k)^{-1}$, which would be 
much larger than the cutoff wavelength and hence compatible with the cutoff.

The IR
contribution (\ref{IRcontribution}) to the total entropy does not 
appear to have any contribution corresponding to momentum 
entanglement, and it should not, since that is ``internal" entanglement
that does not contribute to the total entropy. Therefore,  (\ref{IRcontribution}) must 
differ from the von Neumann entropy
of the reduced density matrix of the lower modes outside the horizon.
We now attempt to identify the origin of this discrepancy
using a formal computation in which the issues raised in
the previous two sections are ignored. This strategy is sensible
because the issue of momentum entanglement is orthogonal 
to the others.

Let $Z = Tr_{a,A} \exp(-\beta H)$ denote the full partition function of the
exterior degrees of freedom, where $a$ and $A$ stand for IR and UV 
degrees of freedom, schematically. If we first trace only over $A$,
we have 
\begin{equation}\label{TrA}
Tr_{A} \exp(-\beta H)= Z_g \exp(-\beta H_a)\,.
\end{equation}
Here $Z_g$ is a $\beta$ dependent number, independent of the 
fields $a$, and $H_a$ is an effective Hamiltonian for the lower modes.
This split may be ambiguous in general, but for the scalar field we
defined $Z_g=\exp(-\Gamma_k[g])$ via the part of the effective action that was independent
of the scalar field, and the remaining effective action was $\tilde{S}_k$,
which is the action that would correspond to $H_a$. In the
presence of interactions we expect $\tilde{S}_k$ to include non-local terms, 
and we expect $\beta$ dependence in $H_a$ simply because it
is defined by a $\beta$ dependent procedure. 

It follows then that $Z = Z_g Z_a$, where $Z_a = Tr_a \exp(-\beta H_a)$.
Now when we compute the contribution to the entropy, 
$-(\beta\partial_\beta - 1)(\ln Z_g + \ln Z_a)$, the $Z_g$ term
contributes a ``gravitational entropy" and the $Z_a$ contribution
corresponds to (\ref{IRcontribution}). If $H_a$ did not depend on
$\beta$ (for example for a non-interacting field), then the 
$Z_a$ term would contribute the von Neumann entropy of the 
density matrix $\rho_a= Z_a^{-1}\exp(-\beta H_a)$. This would
just be the entropy of the $a$ subsystem. However the $\beta$ dependence of 
$H_a$ produces an extra term, 
$\beta^2 \langle \partial_\beta H_a\rangle$. This term
may be the origin of the discrepancy. If it were to contain the 
negative of the momentum entanglement entropy,
it would cancel the contribution of the latter to the von Neumann entropy
term, leaving us with no momentum entanglement in (\ref{IRcontribution}).
We leave a more complete understanding of this point for future work.

\subsection{Non-locality of the effective action} 

One further conceptual issue arising for interacting fields  is that the Wilsonian effective action for the lower $\varphi$ modes, $\tilde{S}_k$, is in general nonlocal 
(though the nonlocality should be suppressed at length scales much longer than $k^{-1}$).
This raises further questions for the interpretation of the IR contribution (\ref{IRcontribution}). 
The Hamiltonian formalism corresponding to a nonlocal action is at least nonstandard, so the
canonical thermal ensemble is nonstandard, and therefore
the relation between the path integral with a nonlocal action and the thermal partition function is unclear.
(The nonlocality problem is less severe at the level of the gravitational effective action $\Gamma_k$, 
where a curvature expansion has the first nonlocalities appearing at order $R^2$ \cite{codello1}, beyond the range of our approximation.) 
We leave clarification of this issue also to future work.

Finally, it is worth mentioning that in the setting of the analysis of Ref.~\cite{onmodel}, the entropy contributions were studied at scales above or below 
mass scales in the theory. Presumably in that setting the presence of the 
mass threshold suppressed any non-locality in the Wilsonian effective action,
in addition to any $\beta$ dependence of the Hamiltonian
analogous to $H_a$ (\ref{TrA}).

\section{Summary and Discussion}

The
aim
of this paper has been to investigate
black hole entropy within the framework provided by the renormalization group, 
in order to probe the role of entanglement entropy in a setting where its contribution
is inherently finite. The idea was to avoid the regulator dependence that arises 
for an otherwise divergent quantity.
In Section 2, we first reviewed how the entropy of the canonical ensemble containing a spherical black hole
is computed (neglecting metric fluctuations) from the on-shell evaluation of the full effective action for the gravitational field, $\Gamma[g]$. 
We then reviewed how the contribution from a minimally coupled matter field
is formally equal to its (divergent) entanglement entropy.

In Section 3, we 
introduced an RG cutoff scale $k$, and defined a flowing effective action $\Gamma_k[g]$ for the metric, 
which excludes the effects of IR excitations that are below the scale $k$. 
The Bekenstein-Hawking entropy computed from this effective action, in terms of the running coupling $G_k$, is complemented by
a contribution from 
the remaining, unintegrated modes to give the total  
entropy, according to (\ref{entropies}) in general, and (\ref{freephiren}) for the massless scalar.
In Section 4, we  
developed a similar decomposition of the entropy
in the case of interacting quantum fields. The upshot is equation (\ref{entropiesint}), which differs from (\ref{entropies}) in that
the quantity $W_k$ encapsulates the contribution of the lower modes not by an explicit one-loop determinant (\ref{actions}), but implicitly through a path integral involving the Wilsonian effective action for the low-energy modes, $\tilde{S}_k[g,\varphi]$ (\ref{Wint}).
The combination $S_k = \Gamma_k+\tilde{S}_k$ evolves with $k$ according to 
(\ref{Polch1}), which is a curved space version  of the Polchinski equation.
The results of Sections 3 and 4 can in principle be extended to include gravitational fluctuations, 
using a background field quantization of gravity along 
the lines of Ref.~\cite{reuterbecker}.

Section 5 was devoted to analyzing whether the contribution from the modes below 
$k$ can be identified with their entanglement entropy as a subsystem. 
We identified a number of problems for this interpretation. Some just concern
the precise definition of the entanglement
and are thus perhaps not very significant,
while others may pose a serious challenge to the very notion of an RG scale 
dependent entanglement entropy. Five issues were discussed in Section 5. 
The first three are relevant for both free and interacting fields, 
while the last two arise only in the presence of interactions:
\begin{itemize}
\item[(i)] The well-known presence of a contact term
in the entropy for non-minimally coupled fields means that the contribution of the 
lower modes cannot  reflect only their entanglement (unless the proposal
mentioned in footnote \ref{0modes} is correct). 
However, the contact term is at least isolated from the rest of the contribution. It can be thought of as an quantum correction to the Noether potential at scale $k$, and thus as part of the ``gravitational" entropy at that scale.

\item[(ii)] Our use of  Euclidean momentum cutoff means that the RG scale has no
direct real space interpretation, although this may not be a serious impediment
since a precise Lorentzian correspondence could perhaps be established, or
the RG scale could be implemented in a different fashion. 

\item[(iii)] The fuzziness of the horizon concept, and therefore of the horizon entanglement 
entropy, in the presence of a momentum 
cutoff is a more basic issue. However this could be looked at as a necessary  
ambiguity in the notion of scale dependent entanglement entropy, and not a
fundamental problem with that notion per se.  

\item[(iv)] Interactions of the matter field produce an entanglement between sub and super $k$
modes that is not included in the contribution of the sub $k$ modes to the total entropy of the thermal
ensemble.  Perhaps the momentum entanglement can be isolated by its coupling constant 
dependence. Also, if the coupling constant $\lambda$ is much smaller than unity,
the momentum entanglement could perhaps be suppressed by focusing on a sufficiently small
neighborhood of the horizon while remaining compatible with the momentum cutoff. 

\item[(v)]
After integrating out the super $k$ modes in an interacting theory the effective action
must be non-local, since the dynamics of the sub $k$ modes is not truly autonomous.
While perhaps only on scales shorter than $k^{-1}$,
this non-locality might invalidate the 
precise
link between the sub $k$ partition function on a cone and
the entanglement entropy, since then no standard Hamiltonian for the system exists.
\end{itemize}

Our conclusion is that when the gravitational black hole entropy is derived from 
the Noether charge 
for an effective action
at scale $k$, the finite remaining contribution to the total
entropy from the IR quantum modes below this scale 
has no straightforward
interpretation as 
entropy of entanglement across the horizon. 
However, for free fields this interpretation may be admissible provided that difficulties (ii-iii) can be 
suitably finessed.
For interacting fields the points raised in (iv-v) raise a larger challenge to 
this interpretation. 
In any case, blithe claims involving that interpretation
should be avoided. 

All these concerns, however, are introduced by the attempt to justify the entanglement interpretation 
for the contribution to the entropy of the modes below a finite energy scale (in order to avoid dealing with divergent quantities). 
Even if this interpretation is not fully justified, it could still be that
the \textit{total} black hole entropy originates as entanglement entropy 
in a UV-complete theory of quantum gravity. 
\\
\\
{\bf Note added:} Another perspective on renormalization of entanglement entropy 
was presented in \cite{luty}, which appeared after this work was completed.

\section{Acknowledgements}

We thank William Donnelly, Daniel Litim, Robert Myers, Martin Reuter, and Aron Wall for helpful discussions.
This research was supported in part by 
the National Science Foundation under Grant Nos. PHY-0903572 and PHY11-25915.

\appendix
\section{Polchinski equation on curved backgrounds}

In this appendix we derive the RG flow equation (\ref{Polch1}) that $S_k[g,\phi]$ satisfies. We consider the definition  (\ref{wilsoniandef}) evaluated for two close RG scales $k$ and $k+\Delta k$, subtract the equations and expand for small $\Delta k$. obtaining:
\begin{equation}\label{deltak}
S_{k+\Delta k}[g,\phi]-S_k[g,\phi]= \frac{\Delta k}{2}\, \mathrm{Tr}\, \langle \varphi\cdot\dot{\mathcal{R}}_k\cdot\varphi\rangle_{\phi, k}\,.
\end{equation}
We use a compact notation, with the overdot being a $k$-detivative, writing $F\cdot D=\int dx\,F(x)\,D(x,y)$ for a function $F$ and an operator $D$, and 
\begin{equation}
\langle F \rangle_{\phi,k} \equiv \frac{\int \mathcal{D}\varphi\,F\,\mathrm{e}^{-\frac{1}{2}\int \varphi \left(-\nabla_g^2+\mathcal{R}_k\right)\varphi- S_b[g,\phi+\varphi]}}{\int \mathcal{D}\varphi\,\mathrm{e}^{-\frac{1}{2}\int \varphi \left(-\nabla_g^2+\mathcal{R}_k\right)\varphi- S_b[g,\phi+\varphi]}}\,.
\end{equation}
The right hand side of (\ref{deltak}) can be related to the functional derivatives of $S_k[g,\phi]$ with respect to $\phi$, using the following relation between two-point operators computed by differentiation of  (\ref{wilsoniandef}):

\begin{equation}
\frac{\delta^2S_k}{\delta\phi\delta\phi}-\frac{\delta S_k}{\delta\phi}\frac{\delta S_k}{\delta\phi}=D_k-\langle \varphi\cdot D_k\, \,D_k\cdot\varphi\rangle_{\phi,k}\,.
\end{equation}
where $D_k$ stands for the two-point operator $(-\nabla^2+\mathcal{R}_k)$.
This leads in the limit $\Delta k\rightarrow 0$ to the flow equation:
\begin{align}\label{PolR}
\dot{S}_k[g,\phi]&=\frac{1}{2}\left\{\mathrm{Tr}\left[\frac{\delta S_k}{\delta\phi}\cdot D^{-1}_k\cdot\dot{\mathcal{R}}_k\cdot D^{-1}_k\cdot\frac{\delta S_k}{\delta \phi}\right]-\mathrm{Tr}\left[D_k^{-1}\cdot\dot{\mathcal{R}}_k \cdot D_k^{-1}\cdot\frac{\delta^2S_k}{\delta\phi\delta\phi}\right]\right.\nonumber\\
&+\left.\mathrm{Tr}\left[\dot{\mathcal{R}}_k\cdot D_k^{-1}\right]\right\}
\end{align}
Using (\ref{defP}), this can be rewritten in a more compact way in terms of the low-momentum propagator $P_k$:
\begin{equation}\label{Polch}
\dot{S}_k[g,\phi]=\frac{1}{2}\left\{\frac{\delta S_k}{\delta\phi}\cdot \dot{P}_k\cdot\frac{\delta S_k}{\delta \phi}-\mathrm{Tr}\left[\dot{P}_k\cdot\frac{\delta^2S_k}{\delta\phi\delta\phi}\right]+\mathrm{Tr}\left[{\dot{P}_k}{(-\nabla_g^2+\mathcal{R}_k)}\right]\right\}
\end{equation}
This is the Polchinski equation in a curved-background setting. Standard presentations of the Wilsonian renormalization group \cite{morris, polonyi} are restricted to flat space and include only the first two terms, omitting the third one since it affects only the gravitational effective action and is thus irrelevant in flat space. (The same thing happens for the normalization factor $N_k$). The expression (\ref{PolR}) in terms of the cutoff function $\mathcal{R}_k$ highlights the similarity to our framework for free fields; note in particular that if $S_k[g,\phi]$ does not depend on $\phi$, as happens when the bare action is a free massless field, the first two terms of (\ref{PolR}) vanish and we recover (\ref{ERGE}). (The form of the equation is the same whether the overdot stands for $\partial_k$ or for $k\partial_k$).

\end{document}